\documentstyle[12pt,openbib,psfig]{article}
\def\be{\begin{equation}}
\def\ee{\end{equation}}

\newcommand{\La}{\Lambda}

\newcommand{\ka}{\kappa}
\newcommand{\s}{\Sigma}
\newcommand{\ca}{\Xi}
\newcommand{\g}{\gamma}
\newcommand{\Om}{\Omega}
\newcommand{\r}{\rho}
\newcommand{\f}{\frac}

\newcommand{\ep}{\epsilon}
\begin{document}
\title{\bf Two new sum rules for octet-baryon magnetic moments ($\mu$)
and constraints on QCD sum rules from new experimental
determination of $\mu$-s for the decuplet.} \vskip .5cm
\author{Monika Sinha $^{1,2}$, Ashik Iqubal $^{3}$, Mira Dey  $^{1,4}$,
 \\ and \\ Jishnu Dey $^{5,4}$}
\vspace{.5 cm}
\date{\today }
\maketitle {\it Abstract} :

Recently the $\mu_{\Delta ^{++}}$ was found from a fit to $\pi^+p$
scattering \cite{castro}. This enable us to pinpoint condensate
parameters more precisely in the context of QCD sum rules (QCDSR).

In the octet sector,  the Coleman-Glashow sum rule (CGSR)
\cite{cg} is violated by the experimental $\mu$-s. QCDSR allows us
to write down two sum rules similar to the CGSR, which are obeyed
by the experimental magnetic moments, whereas they rule out a
specific model using the Wilson loop approach and a particular
chiral quark model.

It is amusing to note that the QCDSR allows us to write down the
quark and gluon condensates in terms of measurables like the
$\mu$-s of the nucleons and the $\Sigma^{\pm}$.

 \vskip .5cm

(1) Abdus Salam ICTP, Trieste, Italy, on leave from Dept. of
Physics, Presidency College, Kolkata 700 073, India

(2) CSIR Research fellow, Govt. of India.

(3) Theory Division, TIFR, Mumbai, India.

(4) {\it{ Work supported in part by DST grant no. SP/S2/K03/01,
Govt. of India, \\ permanent address : 1/10 Prince Golam Md. Road,
Kolkata 700 026, India, email : deyjm@giascl01.vsnl.net.in}}.

(5) Abdus Salam ICTP, Trieste, Italy
 on leave from Dept. of Physics,  Maulana Azad College, Kolkata 700 013, India.

\section{Introduction}

    ~~~~Magnetic moments of baryons depend very sensitively on model
parameters. So accurately measured values of baryon magnetic
moments are very useful to constrain the validity of modeling.

In the decuplet sector $\mu_{\Om}$ was measured accurately and
differs from most of the theoretical estimates, thus posing a
challenge to the latter. It was shown \cite{idd} that this can be
explained from QCDSR and QCD condensate parameters are thereby
constrained.

    Recently there has been much experimental and theoretical studies,
seemingly a little isolated, with different groups not conscious
of each other's work. But these studies of magnetic moments can be
correlated to evolve a picture of the QCD vacuum which is very
rich. Correlations between $\mu_B$ should also be interesting to
experimentalists. Thus  for example Kotulla et al. \cite{kotulla},
during their determination of $\mu_{\Delta^{+}}$ should have found
the QCDSR relation $\mu_{\Delta^{++}}~=~ \f{1}{2}
\mu_{\Delta^{+}}$ interesting, in view of the earlier
determination of $\mu_{\Delta^{++}}$ by Castro and Mariano
\cite{castro}. We write down two other sum rules involving octet
baryon $\mu_B$, hoping to stimulate more studies of these objects.

QCDSR enables us to write down the quark and gluon condensates in
terms of the octet magnetic moments, for example $\mu_p$, $\mu_n$
and $\mu_{\Sigma^\pm}$.

 We also find that the magnetic susceptibility  needs to be very
large to fit the determination of the magnetic moment of
$\Delta^{++}$,  made in \cite{castro},  from the most sensitive
observables in radiative $\pi^+p$ scattering.

    Iqubal et al. \cite{idd} used the QCDSR to fit the $\Om ^-$
magnetic moment. $\mu _{\Om^-}$ has been the subject of many
studies \cite{ddt, lee, lee1, Latt, Rqm, CSqm}. The magnetic
moment was unknown, when the large colour Fock approximation paper
\cite{ddt} was published. But on hindsight, the value predicted
there, within the acceptable parameter range, agrees with the
presently determined experimental result \cite{dt}\footnote{The
methods of this calculation are now used for strange star matter
\cite{dbdrs,li}}. The results of Lee \cite{lee} using QCDSR and
those from the lattice calculation \cite {Latt} underestimated it
whereas the light- cone relativistic quark model \cite{Rqm} and
the chiral quark soliton model \cite{CSqm} overestimated it. This
intriguing situation was investigated by looking at the
calculations of Lee using a slightly different point of view
advocated in \cite{ddsr} and it was found that one indeed gets
good agreement with experiment \cite{idd}.

Further, it was  pointed out in \cite{lee,idd} the  $\mu
_{\Delta^{++}}$ depend sensitively on the magnetic susceptibility.
This moment is now obtained in \cite{castro}. They have determined
the $\mu$ of the  $\Delta^{++}$ resonance by using a full
dynamical model which consistently describes the elastic and
radiative $\pi^+ ~p$ scattering data. It also reproduce very well
the total and differential cross-sections for elastic $\pi^+~p$
scattering close to the resonance region. It provides an amplitude
for radiative $\pi^+~p$ scattering that satisfies electro magnetic
gauge invariance when finite width effects of $\Delta^{++}$
resonance are taken into account. From their determination we can
fix the magnetic susceptibility parameter of QCDSR.

    As already mentioned, very recently Kotulla et al. have
investigated the reaction $\g p \rightarrow \pi^0\, \g\prime\, p$.
Through the reaction channel they arrived at the magnetic dipole
moment of the $\Delta^+$ (1232) resonance \cite{kotulla}. Their
measured value is also consistent with QCDSR.

\begin{table}[h]
\caption{The experimental values of magnetic moments in unit of
$\mu_N$} \vskip 1cm
\begin{center}
\begin{tabular}{|c|c|c|c|c|c|c|c|c|}
\hline

$p$&$n$&$\Xi^-$&$\Xi^0$&$\s^+$&$\s^-$&$\Om^-$&$\Delta^+$&$\Delta^{++}$\\
\hline

2.793&-1.913&-0.6507&-1.25&2.458&-1.16&$2.019$&$2.7_{-2.8}^{+2.5}$&$6.14\pm
0.51$\\

\hline
\end{tabular}
\end{center}
\label{exp}
\end{table}

We have summarized the values of experimentally determined
magnetic moments \cite{kotulla,castro,dt} in the Table \ref{exp}.

The Coleman and Glashow sum rule CGSR \cite{cg} is given by

\be
\Delta
CG~=~\mu_p~-~\mu_n~+~\mu_{\s^-}~+~\mu_{\s^+}~+~\mu_{\Xi^0}~-~\mu_{\Xi^-}~=~0
\label{CGS} \ee

Experimental numbers give $\Delta CG~=~0.49~ \mu _N$.

    ~~From the experimental values of octet magnetic moments we can get
the values of the quark and gluon condensates respectively :

\be
a~=~-2\pi^2<\bar qq>~=~\sqrt{-0.4618(\mu_p~+~2~\mu_
n)~-~1.8382(\mu_{\s^+}~+~2~\mu_{\s^-})} \label{msr1}\ee

\be
b~=~<g_s^2G^2>~=~-4.4545(\mu_ p~+~2~\mu_
n)~-~21.2651(\mu_{\s^+}~+~2~\mu_{\s^-}) \label{msr2}\ee

Putting the values of the experimental moments one gets numerical
values  $a = 0.472$ and $b= 1.667$. The former matches with the
value we use, the latter differs in the last figure, we use 1.664.

We have two new sum rules, SR1 and SR2, resulting from the scaling
of the baryonic coupling to its current \cite{ddsr}. These are as
follows:

\be
\Delta
SR1~=~(\mu_p~+~2\mu_n)~+6.7096(\mu_{\s^+}~+~2\mu_{\s^-})~-~3.4484(\mu_{\Xi^-}~-~\mu_{\Xi^0})~+2.1741~=~0
\label{SR1} \ee
\be
\Delta
SR2~=~(\mu_p~+~2\mu_n)~+~4.7738(\mu_{\s^+}~+~2\mu_{\s^-})~-~0.9988(\mu_{\Xi^-}~-~\mu_{\Xi^0})~+0.9781=~0
\label{SR2} \ee

    Using  the experimental values of magnetic moments \cite{dt} the
left hand side of these two sum rules (Eq. \ref{SR1} and Eq. \ref
{SR2}) give $\Delta SR1~=~4.4929\times 10^{-4} \mu _N$ and $\Delta
SR2 ~=~ 5.3175\times 10^{-3} \mu _N$. These sum rules are very
powerful. For example the chiral quark model for octet baryon
magnetic moments of Dahiya and Gupta \cite{hd} becomes
questionable, although it satisfies the $\Delta CG$ while fitting
the experimental moments approximately. The agreement to $\Delta
CG$ obtained in this paper is clearly accidental; the small
departures from the experimental moments cancel for CGSR, but do
not for $\Delta SR1$ and $\Delta SR2$ (see Table \ref{others}).
\begin{table}[ht]
\caption{The values of magnetic moments and sum rules in unit of
$\mu_N$ for \cite{ha} and \cite{hd}. Note that the values of
$\Delta SR$ should be zero.} \vskip 1cm
\begin{center}
\begin{tabular}{|c|c|c|c|c|c|c|c|c|c|}
\hline

$p$&$n$&$\Xi^-$&$\Xi^0$&$\s^+$&$\s^-$&$\Delta SR1$&$\Delta
SR2$&$\Delta CG$&ref\\ \hline

2.744&-1.955&-0.598&-1.278&2.461&-1.069&0.830&0.675&0.489&[16]\\
\hline
2.800&-1.990&-0.560&-1.240&1.430&-1.200&1.148&0.739&0.480&[15]
\\
\hline
\end{tabular}
\end{center}
\label{others}
\end{table}
However it is possible that with more judicious choice of
parameters the chiral quark model may be able to satisfy the new
sum rules given by us.

    The same comments apply to the model of Ha and Durand
\cite{ha} in Table \ref{others}. They fit the $\Delta CG$ fairly
well but their model fails for $\Delta SR1$ and $\Delta SR2$. The
decomposition of the magnetic moments in terms of the  parameters
of Table VI of their paper may perhaps be used effectively to
satisfy the new sum rules.

\section{QCDSR for decuplet $\mu_B$.}

    ~~~~As is widely known, QCDSR is a very powerful tool in revealing a deep
connection between hadron phenomenology and vacuum structure
\cite{svz} via a few condensates like $a$, $b$, related to the
quark (q) and gluon (G) vacuum expectation values. These can be
used for evaluating $\mu_B$ \cite{si,c}, where some new parameters
enter, for example, $\chi$, $\ka$ and $\xi$, defined through the
following equations :
\be
<\bar q\sigma_{\mu\nu}q>_F = e_q \chi <\bar q q> F_{\mu \nu}
\label{eq:2},
\ee
\be
<\bar q g G_{\mu\nu}q>_F = e_q \ka <\bar q q> F_{\mu \nu}
\label{eq:3},
\ee
\be
<\bar q\ep_{\mu\nu\r\g}G^{\r\g}\g_5q>_F = e_q \xi <\bar q q>
F_{\mu \nu} \label{eq:4}. \ee where the F denotes the usual
external electromagnetic field tensor. Lee \cite{lee} very
carefully evaluated the contributions of these operators to the
magnetic moments of the $\Om ^-$ and $\Delta^{++}$, the latter
emerging from the former when the quark mass $m_s$, is put equal
to zero, the parameter $f$ and $\phi$ are put equal to 1 and the
quark charge $e_s = -1/3$ is replaced by $e_u = 2/3$. The
parameter $f$ and $\phi$ measure the ratio of values for quark
condensates and quark spin-condensates with strange and (ud)
quarks.
\be f = \f{<\bar s s>}{<\bar u u>}
\label{eq:5},
\ee
\be
\phi = \f{<\bar s\sigma_{\mu\nu}s>}{<\bar u\sigma_{\mu\nu}u> }
\label{eq:6}
\ee

For the expression for the  $\mu _{\Om^-}$ and $\Delta ^{++}$ sum
rules we refer the expressions derived in Lee \cite{lee} which we
reproduce here for the sake of completeness, in terms of the Borel
parameter M and the intermediate state contribution A  :

\begin{eqnarray}
& &
  {9\over 28} e_s  L^{4/27} E_1 M^4
- {15\over 7} e_s f \phi  m_s \chi a L^{-12/27} E_0 M^2
+ {3\over 56} e_s  b L^{4/27}
- {18\over 7} e_s f m_s a L^{4/27}
\nonumber \\ & &
- {9\over 28} e_s f \phi (2\ka + \xi) m_s a L^{4/27}
- {6\over 7} e_s f^2 \phi \chi a^2 L^{12/27}
- {4\over 7} e_s f^2 \ka_v a^2 L^{28/27} {1\over M^2}
\nonumber \\ & &
- {1\over 14} e_s f^2 \phi (4\ka + \xi) a^2 L^{28/27} {1\over M^2}
+ {1\over 4} e_s f^2 \phi \chi m^2_0 a^2 L^{-2/27} {1\over M^2}
\nonumber \\ & &
- {9\over 28} e_s f m_s m^2_0 a L^{-10/27} {1\over M^2}
+ {1\over 12} e_s f^2 m^2_0 a^2 L^{14/27} {1\over M^4}
\nonumber \\ & &
= \tilde{\lambda}^2_\Om
\left( {\mu_{\scriptscriptstyle \Om}\over M^2} + A \right) e^{-M^2_\Om/M^2}.
\label{omeg_we5}
\end{eqnarray}
Here
\be
E_n(x) = 1 - e^{-x}\sum_
n \f{x^n}{n!},\; x = w_B^2/M_B^2
\ee
where $w_B$ is the continuum, and
\be
L = \f{ln(M^2/\La ^2_{QCD})}{ln(\mu ^2/\La ^2_{QCD})} \ee For
evaluating the magnetic moment we use the above equation and
divide by the equation for the mass sum rule given earlier by Lee
\cite{lee2}. Thus we eliminate the parameter $\lambda_{\Om ^-}$ in
the spirit of \cite{ddsr} and we get an excellent fit to the
resulting numbers in the form $\mu _{\Om^-} + A/M^2$. We find that
the results are not very sensitive to $\ka_v$, the so called
factorization violation parameter, defined through
\be
<\bar u u \bar u u> = \ka_v <\bar uu>^2. \ee Neither are the
results very sensitive to the parameters $\ka$ and $\xi$. We use
the crucial parameters $a$ and $b$ from \cite{ddsr}, since they
must fit the octet baryon moment-differences ($\mu _p - \mu _n$)
and ($\mu _{\s^+} - \mu _{\s^-}$). It was shown in \cite{ddsr}
that by using the empirical scaling of the $\tilde{\lambda}$ with
the $(baryon\; mass)^3$ - these differences depend only of $a$ and
$b$, and one gets $a = 0.475\; GeV^3$ and $b = 1.695 \;GeV^4$. In
this paper we have used slightly different values 0.472 and 1.664
for $a$ and $b$. Further, to fit the difference ($\mu _{\ca^0} -
\mu _{\ca^-}$), $m_s$ was set to be 170 MeV in \cite{ddsr} and we
use this value.

\begin{table}[ht]
\caption{The values of parameters and their corresponding magnetic
moments} \vskip 1cm
\begin{center}
\begin{tabular}{|c|c|c|c|c|c|c|c|}
\hline

$\chi$&&\multicolumn{3}{c|}{$\mu_{\Om^-}$}&\multicolumn{3}{c|}{$\mu_{\Delta^{++}}$}\\
\cline{3-8} &$\xi~=$&-1&-2&-3&-1&-2&-3\\ \hline

11.0&&-1.945&-1.955&-1.966&5.84&5.87&5.90\\
11.1&&-1.956&-1.966&-1.977&5.89&5.92&5.95\\
11.2&&-1.967&-1.977&-1.988&5.94&5.97&5.99\\
11.3&&-1.978&-1.988&-1.998&5.99&6.02&6.05\\
11.4&&-1.988&-1.999&-2.009&6.04&6.07&6.09\\
11.5&&-1.999&-2.010&-2.020&6.09&6.11&6.14\\
11.6&&-2.010&-2.020&-2.031&6.14&6.16&6.19\\
11.7&&-2.021&-2.032&-2.042&6.18&6.21&6.24\\
11.8&&-2.032&-2.042&-2.053&6.23&6.26&6.29\\

\hline
\end{tabular}
\end{center}
\label{para}
\end{table}

    Table \ref{para} shows the dependence of the magnetic moments on
the parameters. Clearly the agreement with experiment is very good
both for $\mu _{\Delta^{++}}$ and $\mu _{\Om^-}$. Obviously the
former does not depend on $f$ and $\phi$. It is found that $\chi
\sim 11 $ is the best choice for the $\mu _{\Delta^{++}}$. For
such a $\chi$ one should take $\phi~\sim~0.35$ and $f~\sim~0.564$
to get the experimental value of $\mu _{\Om^-}\, =\,( - ~2.019 \pm
0.054)\,\mu _N $ \cite{dt}. The $\mu _{\Delta^{++}}$ is known only
approximately, $(6.14 \,\pm \,0.51)\, \mu _N$ \cite{castro} and a
better determination will enable us to pinpoint $\chi$. As such
the experimental determination is very important since it gives us
a very large magnetic susceptibility $\chi$.

    Dahiya and Gupta \cite{dg}, in their paper on decuplet
$\mu_B$, seem to be unaware of the 2001 publication of
\cite{castro} and their fit to $\mu _{\Delta^{++}}$ is poor.

\section{Results and discussion}

~~~~~We find that using the constrained values of the parameters
$a$ and $b$ \cite{ddsr} one can get a good fit to the known
decuplet magnetic moments. The moments may be used to pinpoint (1)
the susceptibility $\chi$, (2) $f$ and (3) $\phi$, the ratio-s of
the condensate and spin condensate for strange and ud quarks.

    For octet magnetic moments two sum rules are written down from
QCDSR [eqn.(\ref{SR1}) and (\ref{SR2})]. These two sum rules are
used to rule out some specific quark model calculation which claim
to have fitted experimental magnetic moments satisfactorily but
are obviously in contradiction with QCDSR. It is just that the sum
rules highlight the discrepancies in the particular combination of
the moments, to point out the inadequacy of the models. We hope
future theoretical models will try to accommodate these new sum
rules in their fitting while the new experimental data will
continue to satisfy them.

    It is interesting that the quark and gluon condensates
can be written out directly in terms of octet magnetic moments
[eqn.(\ref{msr1}) and (\ref{msr2}.)].

    Finally we hope there will be more experimental data on baryon
magnetic moments since it helps us to pinpoint QCD vacuum
properties via QCDSR technique.

\section{Acknowledgements}

Three of the authors (MS, JD and MD) are grateful to Abdus Salam
ICTP, IAEA and UNESCO for hospitality at Trieste, Italy.

\vskip 2cm.


\begin{thebibliography}{}

\bibitem{castro} G. L. Castro, A. Mariano, Phys.Lett. B517 (2002) 339.

\bibitem{cg} S. Coleman, S. L. Glashow, Phys. Rev. Lett 6 (1961) 423.

\bibitem{idd} A. Iqubal, M. Dey, J. Dey, Phys.Lett. B477 (2000) 125.

\bibitem{kotulla} M. Kotulla et al., nucl-ex/0210040.

\bibitem{ddt} J. Dey, M. Dey and J. LeTourneux, Phys. Rev. D34 (1986) 2104.

\bibitem{lee} F. X. Lee, hep-ph/9707411, Phys.Lett. B419 (1998) 14.

\bibitem{lee1} F. X. Lee, hep-ph/9708323, Phys.Rev. D 57 (1998) 1801.

\bibitem{Latt} D. B. Leinweber, T. Draper and R. M. Woloshyn, Phy. Rev. D
46 (1992) 3067.

\bibitem{Rqm}F. Schlumpf, Phys. Rev. D 48 (1993) 4478.

\bibitem{CSqm} H. C. Kim, M. Praszalowicz and K. Goeke, hep-ph/9706531,
Phys. Rev. 57 (1998) 2859.

\bibitem{dt} K. Hagiwara et al., Phys. Rev. D 66 (2002) 010001.

\bibitem{dbdrs}M. Dey, I. Bombaci, J. Dey, S. Ray and B. C. Samanta,
Phys. Lett. B 438 (1998) 123-128; Addendum B 447 (1999) 352
Erratum B 467 (1999) 303-305 ;

\bibitem{li} X. Li, I. Bombaci, M. Dey, J. Dey, E. P. J. van den Heuvel,
Phys. Rev. Lett. 83 (1999) 3776 ; X. Li, S. Ray, J. Dey, M. Dey
and  I. Bombaci , Astrophys. J. Lett. 527 (1999) L51 ; S. Ray, J.
Dey, M. Dey, K. Ray and B. C. Samanta, Astron. \& Astrophys. Lett.
364 (2000) L89; M. Sinha, M. Dey, S. Ray and J. Dey, Mon. Not. R.
Astron. Soc. 337 (2002) 1368; B. Mukhopadhyay, S. Ray, J. Dey and
M. Dey Astrophys. J. Lett. 584 (2003) L83.

\bibitem{ddsr} J. Dey, M. Dey and M. Sinha Roy, Phys. Lett. B 443 (1998) 293.

\bibitem{hd} H. Dahiya, M. Gupta, Phys. Rev. D 66 (2002) 051501-1.

\bibitem{ha} P. Ha  and  L. Durand,  hep-ph/0212381.

\bibitem{svz} M. A. Shifman, A. I Vainshtein and Z. I. Zakharov, Nucl. Phys.
B 147 (1979) 385, 448.

\bibitem{si} B. L. Ioffe and A. V. Smilga, Nucl. Phys. B 232 (1984) 109.
\bibitem{c}C. B. Chiu, J. Pasupathy and S. J. Wilson, Phys. Rev. D 33 (1986)
1961.

\bibitem{lee2} F. X. Lee, hep-ph/9707332, Phys. Rev. C 57 (1997) 322.



\bibitem{dg} H. Dahiya, M. Gupta, hep-ph/0211127.

\end{thebibliography}
\end{document}